\documentclass[journal]{IEEEtran}

\usepackage{graphicx}
\usepackage{amsmath}
\usepackage{amsfonts}
\usepackage{url}
\usepackage{cite}

\begin{document}

\title{On the Convexity of Common Resource Adequacy and Capacity Accreditation Metrics}

\author{Gord Stephen,~\IEEEmembership{Member,~IEEE}
\thanks{G. Stephen is with the Department
of Electrical and Electronic Engineering, Imperial College London, United Kingdom. email: g.stephen@imperial.ac.uk}
\thanks{Manuscript received XXX; revised YYY.}}

\markboth{Peer review pending, September~2026}%
{Shell \MakeLowercase{\textit{et al.}}: Bare Demo of IEEEtran.cls for IEEE Journals}


\maketitle

\begin{abstract}
Many emerging capacity expansion optimization formulations implicitly presume the convexity (or concavity) of power system adequacy risk (or capacity accreditation) metrics. These assumptions have yet to be formally justified.
We present concise mathematical proofs validating the claim for certain metrics (expected unserved energy and effective load carrying capability), but also provide simple invalidating counterexamples for others (loss-of-load expectation and equivalent firm capacity). Formulations built on non-convex (non-concave) metrics structurally undervalue candidate portfolio adequacy, and so cannot claim to produce least-cost expansion plans.
\end{abstract}


\IEEEpeerreviewmaketitle

\section{Introduction}
 
\IEEEPARstart{V}{ariable} renewable generation, energy storage devices, and weather-dependent thermal generator outages are increasingly consequential to the resource adequacy of modern power systems, but their interdependent contributions are not accurately represented in traditional planning reserve margin frameworks. This is a particular issue for capacity expansion models (CEMs), optimization problems which must incorporate some acceptable representation of system adequacy in order to correctly identify cost-effective, but still resource-adequate, candidate portfolios for new resource investments.

These challenges have motivated the development of new CEM formulations that represent adequacy outcomes as high-dimensional nonlinear functions. CEM analysis based on pre-computed multi-dimensional capacity credit surfaces is being adopted in a growing number of state policy processes \cite{cpuc_irpmethods,nys_scopingplanmethods} and local utility regulatory filings \cite{epe_irp}, while others propose to use iteration and cutting planes to adaptively approximate the underlying multidimensional adequacy risk function, bypassing the capacity accreditation process entirely \cite{hart_cuttingplanes}.
These methods are designed to capture important adequacy interaction effects between resources -- but their validity rests entirely on an assumption that the nonlinear functions being approximated are convex (if upper-bounded, e.g. for adequacy risk metrics) or concave (if lower bounded, e.g. for capacity accreditation  metrics). To our knowledge, these assumptions have yet to be formally justified, despite the growing prevalence and policy influence of these approaches.

This work presents mathematical proofs validating these assumptions for certain metrics (expected unserved energy [EUE] and effective load carrying capability [ELCC]), and simple counterexamples invalidating the general claim for others (loss-of-load expectation [LOLE] and equivalent firm capacity [EFC]). These results are notable given the prevalence of LOLE-based capacity credit surfaces in industry practice -- we show that studies using such surfaces should not claim to produce cost-optimal expansion plans in general.

Section \ref{sec:mathbackground} establishes notation and reviews key mathematical concepts. In Section \ref{sec:ccmetrics}, we show that the (quasi)concavity of portfolio capacity accreditation metrics can be guaranteed by the quasiconvexity of their underlying risk metrics. In Section \ref{sec:riskmetrics}, we discuss the convexity (or lack thereof) of the EUE and LOLE risk metrics. Section \ref{sec:conclusion} provides concluding remarks.

\section{Notation and Mathematical Background}
\label{sec:mathbackground}

We consider system-level adequacy risk and capacity accreditation metrics as defined in terms of a resource portfolio vector $\mathbf{x} \in \mathbb{R}_+^n$ denoting the installed capacity of each system resource, demand time series $\mathbf{d} \in \mathbb{R}_+^m$, and an adequacy risk function $f$ such that, for some established risk threshold $r_{\max}$, $f(\mathbf{x}, \mathbf{d}) \leq r_{\max}$ defines a system adequacy criterion.

Here we review some fundamental definitions and results from convex \cite{boyd_convex} and linear \cite{bertsimas_linear} optimization. These results will underpin the proofs in subsequent sections.

\subsubsection{Convex Sets}

A set of points $\mathbb{C} \subseteq \mathbb{R}^n$ is convex if and only if the line segment between any two points in the set remains within the set. That is, for $\mathbf{x}_1, \mathbf{x}_2 \in \mathbb{C}$ and $\theta \in [0, 1]$, $\theta \mathbf{x}_1 + (1-\theta) \mathbf{x}_2 \in \mathbb{C}$.

\subsubsection{Convex and Quasiconvex Functions}

A function $f$ is quasiconvex if and only if the set $\left\{ \mathbf{x} \ | \ f(\mathbf{x}) \leq c \right\}$ is convex for any $c$. An equivalent definition is that $f(\theta \mathbf{x}_1 + (1 - \theta) \mathbf{x}_2) \leq \max\left\{ f(\mathbf{x}_1), f(\mathbf{x}_2) \right\} \ \forall \ \theta \in [0, 1]$.
A quasiconvex function may be negated into a quasiconcave function, for which $\left\{ \mathbf{x} \ | \ f(\mathbf{x}) \geq c \right\}$ is a convex set or equivalently, $f(\theta \mathbf{x}_1 + (1 - \theta) \mathbf{x}_2) \geq \min\left\{ f(\mathbf{x}_1), f(\mathbf{x}_2) \right\} \ \forall \ \theta \in [0, 1]$. Convex (concave) functions are a special subset of quasiconvex (quasiconcave) functions, for which a more stringent criterion holds: a function $f$ is convex (concave) if and only if $f(\theta \mathbf{x}_1 + (1 - \theta) \mathbf{x}_2) \leq (\geq) \ \theta f(\mathbf{x}_1) + (1-\theta) f(\mathbf{x}_2) \ \forall \  \theta \in [0, 1]$.

\subsubsection{Convex Optimization and Linear Programming}

The definition of a convex (concave) function guarantees it will have a global minimum (maximum) value with no other local minima (maxima). This makes it highly conducive to optimization, particularly when decisions are constrained to a convex set, which guarantees a continuous search space. Optimization problems of this form are defined as convex.

A notable example is $ \min \left\{\mathbf{c}^T\mathbf{x}\  |\  \mathbf{A}\mathbf{x} \leq \mathbf{b} \right\}$, the linear program (LP). The optimal objective value $\mathbf{c}^T\mathbf{x}$ of an LP is also a convex function of $\mathbf{b}$.
LPs are particularly flexible since any arbitrary convex (concave) function can be approximated to arbitrary precision as a maximum (minimum) of a set of linear cutting planes: $f(\mathbf{x}) \approx \max_i (\min_i) \left\{\mathbf{a}_i^T \mathbf{x} + b_i \right\}$. This means that, for convex (concave) $f$, a constraint space $\mathbb{C} = \left\{\mathbf{x} \ | \  f(\mathbf{x}) \leq (\geq) \ c\right\}$ can be linearized directly from the linearization of $f$: $ \mathbb{C} \approx \left\{\mathbf{x} \ | \ \mathbf{a}_i^T\mathbf{x} + b_i \leq (\geq) \ c \ \forall \ i\right\}$.

\section{Concavity of Capacity Accreditation Metrics}
\label{sec:ccmetrics}

For the set of portfolios exceeding some accredited capacity threshold to be convex, the capacity metric must be at least quasiconcave in $\mathbf{x}$, or fully concave if the capacity accreditation function is represented as a set of cutting planes. Given our ultimate interest in system-level adequacy outcomes, we consider capacity accreditation of entire portfolios, rather than the marginal or average contributions of individual resources. 

\subsection{Equivalent Firm Capacity (EFC)}

System EFC is defined as the amount of perfectly available electricity supply that could substitute for a given resource portfolio without changing the system's aggregate adequacy risk \cite{epri_metrics}. We claim that, for a risk assessment function $f(\mathbf{x}, \mathbf{d})$ which is strictly decreasing with respect to perfect capacity additions over the range of system load levels, and at least quasiconvex in $\mathbf{x}$, $\text{EFC}_f(\mathbf{x}, \mathbf{d})$ is quasiconcave in $\mathbf{x}$. Consider that by the quasiconvexity of $f$, for $\theta \in [0, 1]$ we have:

\begin{equation}
\label{eq:efc_f_quasiconcave}
    f(\theta \mathbf{x}_1 + (1 - \theta) \mathbf{x}_2, \mathbf{d}) \leq \max\left\{ f(\mathbf{x}_1, \mathbf{d}), f(\mathbf{x}_2, \mathbf{d}) \right\}
\end{equation}

Now consider a function $\tilde{f}(x; \mathbf{d}) = f(x \mathbf{I}_{\text{firm}}, \mathbf{d})$, where $\mathbf{I}_\text{firm}$ is an indicator vector of all zeros except for the the element corresponding to firm capacity, which takes the value 1.  $\tilde{f}$ must also be strictly decreasing and quasiconvex in $x$. Since $\tilde{f}$ is scalar in $x$ and strictly decreasing over the domain of interest, it is invertible, so a decreasing inverse function $x = \tilde{f}^{-1}(r; \mathbf{d})$ must exist which maps any risk level to a unique amount of equivalent firm capacity. Applying this function to (\ref{eq:efc_f_quasiconcave}) yields:

\begin{equation}
\label{eq:efc_quasiconcave}
\begin{split}
    \tilde{f}^{-1}(f(\theta \mathbf{x}_1 + (1 - \theta) \mathbf{x}_2, \mathbf{d}); \mathbf{d}) \geq \\
    \min\left\{ \tilde{f}^{-1}(f(\mathbf{x}_1, \mathbf{d}); \mathbf{d}), \tilde{f}^{-1}(f(\mathbf{x}_2, \mathbf{d}); \mathbf{d}) \right\}
    \end{split}
\end{equation}

\begin{figure*}[!t]
    \centerline{\includegraphics[width=\textwidth]{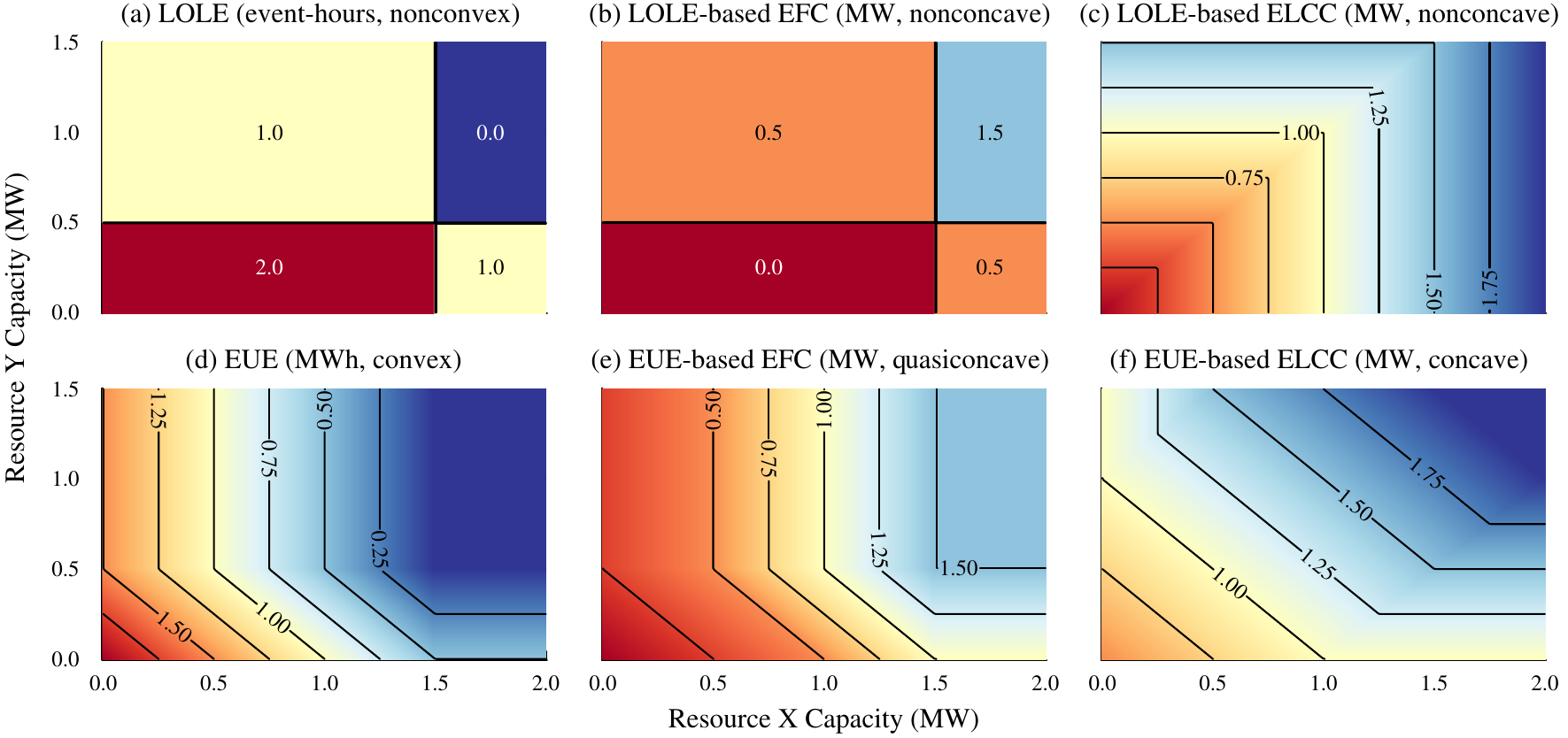}}
    \caption{Example adequacy surfaces for a simple, deterministic, two hourly timestep system with load [0.5, 1.5] MW and two potential capacity resources, Resource X with availability [0\%, 100\%] and Resource Y with availability [100\%, 0\%]. The color scale denotes generalized adequacy levels (higher risk metrics or lower capacity metrics in red, the opposite in blue). Subfigures (a), (b), and (c) (LOLE-based metrics) show nonconvex feasibility regions, e.g for $\text{LOLE} \leq 1 \text{h}$ or EFC and ELCC $\geq 0.5\text {MW}$. Subfigure (e) (EUE-based EFC) shows convex feasibility regions even though the underlying function is nonconcave (note the increase in contour density past the 0.5 MW contour line). Subfigures (d) and (f) (EUE and EUE-based ELCC) depict fully convex/concave functions. (ELCC is calculated here relative to a 1 event-hour LOLE or 1 MWh EUE adequacy criterion.)}
    \label{fig:metric_heatmaps}
\end{figure*}

Since $\text{EFC}_f(\mathbf{x}, \mathbf{d}) = \tilde{f}^{-1}(f(\mathbf{x},\mathbf{d}); \mathbf{d})$, (\ref{eq:efc_quasiconcave}) defines $\text{EFC}_f$ as quasiconcave. Note that even if $f$ is fully convex, $\text{EFC}_f$ is \textit{not} generally concave: Figure \ref{fig:metric_heatmaps}e provides a non-concave counterexample. EFC surfaces are therefore \textit{not} in general representable as the minimum points in a set of cutting planes, and formulations that assume otherwise may underestimate the EFC of candidate portfolios.

\subsection{Effective Load Carrying Capability (ELCC)}

System ELCC is defined as the maximum amount of electricity demand a given resource portfolio is able to serve while remaining below a desired adequacy risk level $r_{\max}$. While this demand could in theory be defined in terms of any time series profile, accepted practice is to consider the prospective demand as a uniform constant $d$ across time \cite{epri_metrics}. We can specify this formally as follows:

\begin{equation}
\label{eq:elcc}
    \text{ELCC}_f(\mathbf{x}, r_{\max}) = \max \left\{ d \ | \ f(\mathbf{x}, d \mathbf{1}) \leq r_{\max}\right\}
\end{equation}

We claim that, for a risk assessment function $f(\mathbf{x}, \mathbf{d})$ which is at least quasiconvex in $\mathbf{x}$ and $\mathbf{d}$, $\text{ELCC}_f(\mathbf{x}, r_{\max})$ is concave in $\mathbf{x}$. Consider two portfolios, $\mathbf{x}_1$ and $\mathbf{x}_2$, and three capacity accreditations, $d_1^* = \text{ELCC}_f(\mathbf{x}_1, r_{\max})$, $d_2^* = \text{ELCC}_f(\mathbf{x}_2, r_{\max})$, and $d_{12}^* = \text{ELCC}_f(\theta \mathbf{x}_1 + (1-\theta)\mathbf{x}_2, r_{\max})$ for some $\theta \in [0, 1]$. Since $d_1^*$ and $d_2^*$ are by construction valid ELCCs, from (\ref{eq:elcc}) we have that $\max \left\{ f(\mathbf{x}_1, d_1^* \mathbf{1}),  f(\mathbf{x}_2, d_2^* \mathbf{1}) \right\} \leq r_{\max}$.

Considering this together with the quasiconvexity of $f$, we see that $f(\theta \mathbf{x}_1 + (1-\theta)\mathbf{x}_2, \left(\theta d_1^* + (1-\theta) d_2^* )\mathbf{1}\right) \leq r_{\max}$, indicating that $\theta d_1^* + (1-\theta) d_2^*$ is a feasible, but not necessarily optimal, choice for $d_{12}^*$. Thus $d_{12}^* \geq \theta d_1^* + (1-\theta) d_2^*$ and, by the definitions of $d_1^*$, $d_2^*$, and $d_{12}^*$, $\text{ELCC}_f$ is concave in $\mathbf{x}$.

\section{Convexity of System Adequacy Metrics}
\label{sec:riskmetrics}

For the set of portfolios considered adequate relative to some maximum risk threshold to be convex, the risk metric must be at least quasiconvex in $\mathbf{x}$, or fully convex if the adequacy risk function is represented as a set of cutting planes.

\subsection{Loss-of-Load Expectation (LOLE)}

LOLE is an expected count of shortfall event-periods (e.g., event-hours or event-days) \cite{epri_metrics}. In the deterministic case it is therefore restricted to whole numbers, and is a stepwise decreasing function of portfolio selection. In the probabilistic case it can take on a continuous range of values, but must still vary stepwise when calculated by averaging over a finite set of probabilistic outcomes (each of which remains a whole number), as is the case in the standard approach of estimating it via Monte Carlo sampling. As a stepwise function, LOLE is nonconvex. It is also not universally quasiconvex in $\mathbf{x}$ -- Figure \ref{fig:metric_heatmaps}a provides a non-quasiconvex counterexample. The lack of such a guarantee means that LOLE-based adequacy criteria are not generally appropriate for formulating convex feasibility sets, whether directly as a maximum risk metric constraint or indirectly as the basis for a minimum EFC or ELCC constraint. Erroneous convexity assumptions about the feasible design space over-constrain the problem and spuriously reject otherwise-valid (and potentially lower-cost) portfolio options.

\subsection{Expected Unserved Energy (EUE)}

EUE is a continuous measure of the expected total energy magnitude of electricity supply shortfalls \cite{epri_metrics}. Multiple levels of shortfall may be possible for a single portfolio $\mathbf{x}$ and demand profile $\mathbf{d}$ (for example, under alternate storage charge and discharge policies), so EUE cannot be uniquely specified from these inputs alone. Since the fundamental mandate of a power system operator is to avoid unserved energy, one could reasonably define $\text{EUE}(\mathbf{x}, \mathbf{d})$ as averaging the \textit{lowest} shortfall level possible among the valid dispatches $\mathbf{y} \in Y_\zeta(\mathbf{x}, \mathbf{d})$ given a random variable $\mathbf{\zeta}$ representing operating uncertainties:

\begin{equation}
    \text{EUE}(\mathbf{x}, \mathbf{d}) = \text{E}[\min \{\text{shortfall}(\textbf{y}) | \mathbf{y} \in Y_\zeta(\mathbf{x}, \mathbf{d}) \}]
\end{equation}

When unserved energy is positive, there exists at least one period in which shortfall occurs and could be reduced with additional firm capacity. Unserved energy is therefore strictly decreasing with firm capacity, for all points with positive minimum unserved energy, so $\text{EUE}(\mathbf{x}, \mathbf{d})$ must be as well.

Furthermore, if these minimizations are expressible as LPs of the form $\min \left\{ \mathbf{c}^T\mathbf{y} \ | \ \mathbf{A}_\mathbf{\zeta} \mathbf{y} \leq \mathbf{b}_{0\zeta} + \mathbf{G}_\zeta\mathbf{x} - \mathbf{H}_\zeta\mathbf{d}\right\}$ (as effectively all linear adequacy dispatch formulations are), the minimum shortfall is a convex function of $\mathbf{x}$ and $\mathbf{d}$, so $\text{EUE}(\mathbf{x}, \mathbf{d})$ is also convex.

This convexity guarantee means that EUE (as formulated here) can be used to define convex feasibility sets, either directly as a maximum risk metric constraint or as the underlying metric for a minimum ELCC constraint. The strictly-decreasing property means that it can be used as the basis for a convex minimum EFC set as well.

\section{Conclusion}
\label{sec:conclusion}

This work provides a mathematical grounding for emerging adequacy constraint formulations in capacity expansion problems. We justify the key assumption underpinning the direct EUE function proposed in \cite{hart_cuttingplanes}, and show that an EUE-based system ELCC surface is representable in a linear optimization.

However, we raise questions about the general validity of the other constraint designs considered here. Any LOLE-derived surface is potentially problematic, and while the portfolios meeting some minimum EFC form a convex set, which could in theory be used to constrain a convex optimization, EFC is not in general a concave function. It is therefore incorrect to assume it can be represented as the minimum of a set of linear cutting planes, as is common practice today.

These results do not \textit{necessarily} imply that any analyses using LOLE-based capacity credit surfaces  (including \cite{epe_irp, cpuc_irpmethods, nys_scopingplanmethods}) are fundamentally flawed -- but any cost-optimality claims will need to be proven by some other means, as the chosen adequacy constraint formulation provides no such guarantee.

\section*{Acknowledgment}
Special thanks to Sleiman Mhanna for discussions which prompted this work, and to Mark O'Malley, Genevieve de Mijolla, and Tom Acker for feedback on early drafts. This work was supported by the U.K. Engineering and Physical Sciences Research Council grant number EP/Y025946/1 and by the Leverhulme International Professorship grant reference LIP-2020-002.


\bibliographystyle{IEEEtran}
\bibliography{references}


\end{document}